\documentclass[preprint,amsmath,amssymb,prl]{revtex4}
\usepackage{graphicx}

\def\de{\delta}

\def\th{\theta}

\def\vs{\varsigma}

\def\ph{\phi}

\def\ps{\psi}

\def\De{\Delta}

\def\Ph{\Phi}
\def\Ps{\Psi}

\def\vev#1{\langle {#1}\rangle}

\def\fr#1#2{{{#1} \over {#2}}}
\def\frac#1#2{{{#1} \over {#2}}}
\def\frac#1#2{{\textstyle{{#1}\over {#2}}}}

\def\lsim{\mathrel{\rlap{\lower4pt\hbox{\hskip1pt$\sim$}}
    \raise1pt\hbox{$<$}}}
\def\gsim{\mathrel{\rlap{\lower4pt\hbox{\hskip1pt$\sim$}}
    \raise1pt\hbox{$>$}}}
\def\sqr#1#2{{\vcenter{\vbox{\hrule height.#2pt
         \hbox{\vrule width.#2pt height#1pt \kern#1pt
         \vrule width.#2pt}
         \hrule height.#2pt}}}}
\newcommand{\beq}{\begin{equation}}
\newcommand{\eeq}{\end{equation}}
\newcommand{\bea}{\begin{eqnarray}}
\newcommand{\eea}{\end{eqnarray}}
\newcommand{\rf}[1]{(\ref{#1})}

\def\syjm#1#2{\phantom{}_{#1}Y_{#2}}
\def\etal{{\it et al.}}

\def\kjm#1#2#3{k^{(#1)}_{(#2)#3}}
\def\cjm#1#2#3{c^{(#1)}_{(#2)#3}}
\def\kI{\cjm{d}{I}{jm}}
\def\kE{\kjm{d}{E}{jm}}
\def\kB{\kjm{d}{B}{jm}}
\def\kV{\kjm{d}{V}{jm}}
\def\kIdjm#1#2{\cjm{#1}{I}{#2}}
\def\kEdjm#1#2{\kjm{#1}{E}{#2}}
\def\kBdjm#1#2{\kjm{#1}{B}{#2}}
\def\kVdjm#1#2{\kjm{#1}{V}{#2}}

\begin{document}

\title{Constraints on relativity violations from gamma-ray bursts}
\author{V.\ Alan Kosteleck\'y$^a$ and Matthew Mewes$^b$}

\affiliation{$^a$Physics Department, Indiana University, 
Bloomington, IN 47405, USA\\
$^b$Physics Department, Swarthmore College,
Swarthmore, PA 19081, USA}

\date{IUHET 572, January 2013}

\begin{abstract}

Tiny violations of the Lorentz symmetry of relativity 
and the associated discrete CPT symmetry 
could emerge in a consistent theory of quantum gravity such as string theory.
Recent evidence for linear polarization in gamma-ray bursts
improves existing sensitivities to Lorentz and CPT violation 
involving photons by factors ranging from ten to a million. 

\end{abstract}

\maketitle

Observations of photon behavior
provide crucial probes of fundamental physics.
Famous examples include 
the classic Michelson-Morley, Kennedy-Thorndike, 
and Ives-Stilwell experiments
\cite{mmktis},
which support the foundational Lorentz invariance of relativity.
In recent years,
a wide range of astrophysical, solar-system, and laboratory tests 
of Lorentz symmetry and its associated discrete CPT symmetry 
have achieved impressive sensitivities using photons
(see Ref.\ \cite{tables} for a compilation).
One motivation for these efforts is the prospect 
that tiny violations of these invariances 
could emerge in a consistent theory of quantum gravity such as string theory 
\cite{ksp}.
In this paper,
we use recent measurements of linear polarization in light
from gamma-ray bursts (GRB)
\cite{mcglynn,dy1,dy2}
to improve existing sensitivities 
to a variety of Lorentz- and CPT-violating effects
by factors ranging from ten to a million.
The key to the exceptional GRB sensitivity to Lorentz and CPT violation 
lies primarily in the extreme propagation distances
during which tiny effects can accumulate,
along with the comparatively high photon energies involved.

At accessible energies,
violations of Lorentz invariance 
are governed by the Standard-Model Extension (SME)
\cite{sme},
a comprehensive effective field theory
containing both General Relativity and the Standard Model
that provides a general theoretical framework for observational studies.
This theory also describes CPT violation in the context
of realistic field theory
\cite{owg}.
The SME action is a sum of coordinate-invariant terms,
including ones formed from Lorentz-violating operators
contracted with controlling coefficients,
and the mass dimension $d$ of each operator
fixes the dimensionality of the corresponding coefficient
\cite{kp}.
In the photon sector,
all gauge-invariant operators describing
the propagation of light have been classified and enumerated
for arbitrary $d$
\cite{km09}. 

The SME predicts that light propagates 
in the presence of Lorentz and CPT violation 
as the superposition 
of two normal modes
that may differ in speed and polarization.
The dispersion relations connecting the photon energy $E$ and momentum $p$ 
for the two modes can be written in the compact but implicit form 
\cite{sme,km09}
\beq
E = \big(1-\vs^0 
\pm \sqrt{(\vs^1)^2+(\vs^2)^2+(\vs^3)^2}\, \big) p ,
\label{dr}
\eeq
where the dimensionless quantities $\vs^a = \vs^a(E,\th,\ph)$ 
depend both on $E$
and on the photon direction of propagation,
which for an astrophysical point source
is fixed by the source codeclination 
$\th \equiv (90^\circ - \text{declination})$
and by the right ascension $\ph$.
For light propagating {\it in vacuo}, 
the quantities $\vs^a$ are linear combinations
of the basic SME coefficients 
$\kI$, $\kE$, $\kB$, and $\kV$,
where $j$, $m$ are angular quantum numbers. 
For each $d$,
many different SME coefficients control the behavior of light.
A given point source at a fixed sky location 
can therefore access only a limited number of coefficient combinations.
Consequently,
multiple sources at different sky locations are required
to constrain fully the coefficient space.

Two major features,
dispersion and birefringence,
can be exploited
to search for Lorentz violation
in radiation from sources at 
cosmological distances.
Dispersion is a characteristic of all SME operators with $d\neq 4$.
Arrival-time differences
in high-energy photons from sources such as GRB 
can be used to constrain the energy dependence 
in the group velocity.
The SME framework shows that dispersion for operators with odd $d$ 
is necessarily accompanied by birefringence,
implying different speeds for the two normal modes.
As a result,
a wave packet not only disperses
but gradually splits into two.
Only the CPT-even operators with even $d$ 
characterized by the coefficients $\kI$ 
give dispersion without birefringence.
GRB constraints on Lorentz-violating dispersion for even dimensions $d$ 
have been obtained for $d=6$ and $8$
\cite{kmapjl,km09,fermi,fermi2}.

Birefringence studies of astrophysical sources such as GRB
offer extreme sensitivity to Lorentz and CPT violation.
The primary signature of birefringence
is a change in photon polarization due to propagation.
This is governed by the phase difference of the eigenmodes
developed during propagation,
which increases with energy.
While dispersion can uniquely constrain the coefficients $\kI$,
for $d\geq 4$ 
birefringence at GRB energies typically offers 
many orders of magnitude better sensitivity 
to the coefficients $\kE$, $\kB$, $\kV$ 
because a comparable dispersion test
would require a time resolution comparable
to the tiny inverse photon frequency.

\begin{figure}
\includegraphics[width=0.4\textwidth]{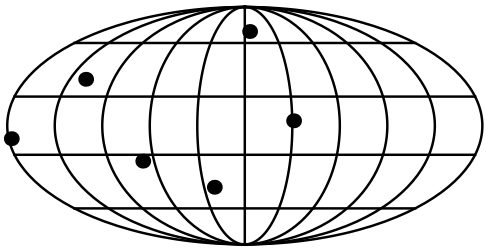}
\caption{\label{skymap}
Sky map showing the locations of the six GRB listed in Table \ref{sources}.
The map is centered at $(\th,\ph) = (90^\circ,0^\circ)$.}
\end{figure}

In one study,
evidence for polarization at the level of
$\Pi > 35\%$ in GRB 930131 and
$\Pi > 50\%$ in GRB 960924
was extracted from data obtained by
the Burst and Transient Source Experiment (BATSE)
\cite{willis}.
These results have been used to
constrain Lorentz violation
for $d = 4,5,6,7,8,9$ \cite{km06,km09}.
Polarization as high as $\Pi = 96^{+39}_{-40}\%$
was identified in GRB 041219A 
using instruments aboard the
International Gamma-Ray Astrophysics Laboratory (INTEGRAL)
\cite{mcglynn}.
This result was used by Stecker
to place bounds on $d=5$ coefficients
on the order of $10^{-34}$ GeV$^{-1}$ 
\cite{stecker}.
Another INTEGRAL analysis found
similarly high degrees of polarization in GRB 041219A,
placing bounds on the single isotropic $d=5$ coefficient 
\cite{laurent}.
A recent analysis using polarization data 
for GRB 100826A ($>6\%$),
GRB 110301A ($>31\%$), 
and GRB 110721A ($>35\%$)
from the Gamma-ray Burst Polarimeter (GAP)
on the Interplanetary Kite-craft Accelerated by Radiation of the Sun (IKAROS)
\cite{dy1,dy2}
also bounded the single isotropic $d=5$ coefficient 
\cite{toma}.
The basic features of all six GRB are summarized in Table \ref{sources},
and their celestial locations are displayed in Fig.\ \ref{skymap}.

\begin{table}
\begin{tabular}{c||c|c|c}
  & $z$ & energy & $(\th,\ph)$ \\
\hline\hline
GRB 930131 & 0.1 \cite{km06} & 31 -- 98 keV \cite{willis} & $(98^\circ,182^\circ)$ \cite{batse}\\
GRB 960924 & 0.1 \cite{km06} & 31 -- 98 keV \cite{willis} & $(87^\circ,37^\circ)$ \cite{batse}\\
\hline
GRB 041219A & 0.02 \cite{laurent} & 100-- 1000 keV \cite{mcglynn} & $(27^\circ,6^\circ)$ \cite{041219}\\
GRB 100826A & 0.71 \cite{toma} & 70 -- 300 keV \cite{dy1} & $(112^\circ,279^\circ)$ \cite{100826}\\
GRB 110301A & 0.21 \cite{toma} & 70 -- 300 keV \cite{dy2} & $(61^\circ,229^\circ)$ \cite{110301}\\
GRB 110721A & 0.45 \cite{toma} & 70 -- 300 keV \cite{dy2} & $(129^\circ,333^\circ)$ \cite{110721}
\end{tabular}
\caption{\label{sources}
GRB for which strong evidence of linear polarization exists.
The second column gives the estimated lower limit on the red shift.
The third column is the energy range in which polarization is observed.
The last column gives the GRB codeclination $\th$ and right ascension $\ph$.
The first two GRB were previously studied \cite{km06,km09},
while the others are the subject of the present work.}
\end{table}

In this work,
we use the polarization reported 
for the four latest GRB to
place improved bounds on direction-dependent combinations 
of SME coefficients for $d=4,5,6,7,8,9$.
For the analysis,
it is useful to expand the quantities $\vs^a$ in energy $E$
\cite{kmapjl,km09},
\beq
\vs^a(E,\th,\ph) 
= \sum_d E^{d-4}\, \vs^{(d)a}(\th,\ph),
\quad (a=0,1,2,3) ,
\eeq
where $\vs^{(d)a}(\th,\ph)$
are direction-dependent combinations of SME coefficients.
The direction dependence can be displayed explicitly 
by further expansion using conventional spherical harmonics 
$\syjm{}{jm}(\th,\ph)$ for $\vs^{(d)0}(\th,\ph)$ and $\vs^{(d)3}(\th,\ph)$,
\bea
\vs^{(d)0}(\th,\ph) &=& \sum_{jm} \syjm{}{jm}(\th,\ph)\, \kI ,
\notag \\
\vs^{(d)3}(\th,\ph) &=& \sum_{jm} \syjm{}{jm}(\th,\ph)\, \kV ,
\label{vac_exp1}
\eea
and their cousins $\syjm{\pm 2}{jm}(\th,\ph)$ of spin-weight two
for the combinations 
$\vs^{(d)\pm} \equiv \vs^{(d)1} \mp i \vs^{(d)2}$,
\beq
\vs^{(d)\pm}(\th,\ph) = 
\sum_{jm} \syjm{\pm2}{jm}(\th,\ph) \big(\kE \pm i\kB\big) .
\label{vac_exp2}
\eeq
The basic SME coefficients $\kI$, $\kE$, and $\kB$ characterize
CPT-preserving Lorentz violation,
while $\kV$ also controls CPT violation.
Table \ref{coefs} summarizes some properties of these coefficients.

\begin{table}
\begin{tabular}{l|c|c|c|c|c}
  & coeff.\ & CPT & $d$ & $j$ & biref.\\
\hline\hline
general
&$\kI$ & $+$ & $4,6,8,\ldots$ & $0,1,\ldots,d-2$ & \\
&$\kE$ & $+$ & $4,6,8,\ldots$ &$2,3,\ldots,d-2$ & $\checkmark$\\
&$\kB$ & $+$ & $4,6,8,\ldots$ &$2,3,\ldots,d-2$ & $\checkmark$\\
&$\kV$ & $-$ & $3,5,7,\ldots$ &$0,1,\ldots,d-2$ & $\checkmark$\\
\hline
isotropic
&$\kIdjm{d}{00}$ & $+$ & $4,6,8,\ldots$ &$0$ & \\
&$\kVdjm{d}{00}$ & $-$ & $3,5,7,\ldots$ &$0$ & $\checkmark$
\end{tabular}
\caption{\label{coefs}
Index ranges and properties of coefficients for Lorentz and CPT violation,
where $-j\leq m\leq j$ as usual.
The isotropic limit is shown in the last two lines.}
\end{table}

Table \ref{coefs} also contains information
about the isotropic coefficients for which $j=m=0$,
which represent a popular restriction of the general framework.
In this limit, 
each value of $d$ has exactly one SME coefficient, 
which is nonbirefringent in the CPT-even case
and birefringent in the CPT-odd case.
The theoretical motivation for this restriction 
is open to doubt because
isotropy can hold only in a single inertial frame,
which cannot be an Earth-based frame 
and requires fine tuning to match the standard Sun-centered frame.
However,
the isotropic limit does offer an order-of-magnitude measure 
of the reach for a given source,
and it also simplifies many equations.
For example,
the general expression for the defect 
in the group velocity relevant for dispersion studies can be written
$\de v_{gr} = (d-3)E^{d-4} \big(-\vs^{(d)0} \pm \big|\vs^{(d)+} \big| \big)$
in the CPT-even case and
$\de v_{gr} = \pm (d-3)E^{d-4} \big|\vs^{(d)3} \big|$
in the CPT-odd case,
but in the isotropic limit one obtains instead the simpler expressions 
$\de v_{gr} = -(d-3)E^{d-4}\kIdjm{d}{00}/\sqrt{4\pi}$
in the CPT-even case and
$\de v_{gr} = \pm(d-3)E^{d-4}\kVdjm{d}{00}/\sqrt{4\pi}$
in the CPT-odd case.
Note that Table \ref{coefs} reveals 
there are no $j=0$ isotropic coefficients
$\kE$ or $\kB$,
so the quantities $\vs^{(d)\pm}$
are necessarily direction dependent.

During birefringent propagation of light,
the difference in phase speed between the two normal modes 
generates a relative phase shift
and a corresponding change in the net polarization.
This change can be visualized 
as a rotation of the Stokes vector 
$\vec s = (s^1,s^2,s^3) = (Q,U,V)$
about a rotation axis 
$\vec\vs = (\vs^1,\vs^2,\vs^3)$
by an angle $\Ph$ equal to the total relative phase shift
\cite{km}.
If $\vec s$ lies along $\vec\vs$,
then the light is in one of the two normal modes
and the polarization remains unchanged.
Note that in the CPT-odd case
the two normal modes are circularly polarized, 
$\vec \vs=(0,0,\vs^3)$,
while in the CPT-even case
they are linearly polarized, 
$\vec \vs=(\vs^1,\vs^2,0)$.
Also,
isotropic birefringence occurs only in the CPT-odd case
where $\vs^3 = E^{d-4} \kVdjm{d}{00}/\sqrt{4\pi}$ 
and for which $d$ is odd,
so changes in polarization for even $d$
necessarily depend on the source position.

For given $d$,
the rotation angle $\Ph$ depends on the difference in phase speed,
$\De v = 2 E^{d-3} |\vs^{(d)a}|$,
where $a=3$ in the CPT-odd case and $a=+$ in the CPT-even case.
It can be found by integrating the accumulated phase
over the propagation time,
$\Ph = \int E \De v\, dt$.
Incorporating the redshift gives 
\beq
\Ph = 2E^{d-3} L^{(d)} \big|\vs^{(d)a}(\th,\ph)\big| ,
\quad a = (3, +),
\label{phi}
\eeq
where $L^{(d)} = \int_0^z (1+z)^{d-4} H_z^{-1}\, dz$
is the effective baseline for dimension $d$ 
in terms of the source redshift $z$ 
and the Hubble expansion rate $H_z$ at redshift $z$.

For the CPT-odd case with $a=3$,
the Stokes vector rotates about the $s^3$ axis,
shifting the linear polarization angle $\ps$ by $\de \ps = \Ph/2$.
A detailed analysis searching for CPT violation
could take advantage of the $E^{d-3}$ dependence in Eq.\ \rf{phi}.
For example,
assuming two photons of energies $E_1$ and $E_2$
initially have the same polarization,
the difference $\ps_2 - \ps_1$ in their polarization angles 
after traveling the effective baseline $L^{(d)}$ 
is given by
\beq
\fr{\ps_2 - \ps_1}
{(E_2^{d-3}-E_1^{d-3})L^{(d)}}
=\sum_{jm} \syjm{}{jm}(\th,\ph)\, \kV .
\eeq
Using observations of GRB 041219A
and assuming only the single isotropic coefficient with $d=5$
\cite{mp},
a search of this type was performed
in Ref.\ \cite{laurent},
yielding the bound 
$\big|\kVdjm{5}{00}\big| < 3\times 10^{-33} \text{ GeV}^{-1}$.

An alternative approach offering conservative constraints 
is to seek a significant degree of linear polarization 
within a given energy band.
Differential rotations within the band smear the polarization
and hence decrease the effective degree of linear polarization,
$\Pi_\text{eff} = \sqrt{\vev{s^1}^2 + \vev{s^2}^2}$.
In the CPT-odd case,
this smearing produces an upper bound in the measured polarization
$\Pi \leq \sqrt{\vev{\cos\Ph}^2+\vev{\sin\Ph}^2}$,
where the equality holds for an initial
100\% linear polarization at constant angle.
The polarization smearing will be nearly complete 
unless the change in $\Ph$ over the energy band is less than $2\pi$.
Under the assumption of only the single isotropic coefficient with $d=5$,
a similar idea has been applied to observations 
of GRB 100826A, GRB 110301A, and GRB 110721A
\cite{toma},
producing the bound
$\big|\kVdjm{5}{00}\big| < 6\times 10^{-34} \text{ GeV}^{-1}$.

In the general case,
for any given odd $d$ and including all coefficients for CPT violation,
we obtain the conservative limit
\beq
\bigg|\sum_{jm} \syjm{}{jm}(\th,\ph)\, \kV \bigg| <
\fr{\pi}{\big|E_2^{d-3}-E_1^{d-3}\big| L^{(d)}} ,
\eeq
where $E_1$ and $E_2$ are the edges of the energy band.
This expression can be used to obtain substantially improved sensitivities 
to CPT violation from GRB 041219A, GRB 100826A, GRB 110301A, and GRB 110721A.
We consider CPT-violating operators of dimensions $d=5$, $7$, and $9$,
for which there are 16, 36, and 64 independent vacuum coefficients,
respectively
\cite{km09}.
Each source generates a new constraint
on a different combination of coefficients,
listed in Table \ref{bounds}.
For completeness,
we also list bounds in the isotropic limit.
These results represent improvements in sensitivity
over existing limits on violations of CPT symmetry
ranging from tenfold to about a millionfold.

\begin{table*}
\scriptsize
\begin{tabular}{c||c|c|c|c}
&
GRB 041219A & 
GRB 100826A & 
GRB 110301A &
GRB 110721A \\
&
$(27^\circ,6^\circ)$ &
$(112^\circ,279^\circ)$ &
$(61^\circ,229^\circ)$ &
$(129^\circ,333^\circ)$ \\
\hline\hline
$\big| \sum_{jm} \syjm{}{jm} (\th,\ph) ~\kVdjm{5}{jm} \big|$ &
$<2\times 10^{-34}$ GeV$^{-1}$ &
$<7\times 10^{-35}$ GeV$^{-1}$ &
$<2\times 10^{-34}$ GeV$^{-1}$ &
$<1\times 10^{-34}$ GeV$^{-1}$ 
\\
$\big| \sum_{jm} \syjm{}{jm} (\th,\ph) ~\kVdjm{7}{jm} \big|$ &
$<2\times 10^{-28}$ GeV$^{-3}$ &
$<4\times 10^{-28}$ GeV$^{-3}$ &
$<2\times 10^{-27}$ GeV$^{-3}$ &
$<8\times 10^{-28}$ GeV$^{-3}$
\\
$\big| \sum_{jm} \syjm{}{jm} (\th,\ph) ~\kVdjm{9}{jm} \big|$ &
$<2\times 10^{-22}$ GeV$^{-5}$ &
$<2\times 10^{-21}$ GeV$^{-5}$ &
$<2\times 10^{-20}$ GeV$^{-5}$ &
$<5\times 10^{-21}$ GeV$^{-5}$
\\
\hline
$\big| \kVdjm{5}{00} \big|$ &
$<8\times 10^{-34}$ GeV$^{-1}$ &
$<2\times 10^{-34}$ GeV$^{-1}$ &
$<9\times 10^{-34}$ GeV$^{-1}$ &
$<4\times 10^{-34}$ GeV$^{-1}$
\\
$\big| \kVdjm{7}{00} \big|$ &
$<8\times 10^{-28}$ GeV$^{-3}$ &
$<1\times 10^{-27}$ GeV$^{-3}$ &
$<8\times 10^{-27}$ GeV$^{-3}$ &
$<3\times 10^{-27}$ GeV$^{-3}$ 
\\
$\big| \kVdjm{9}{00} \big|$ &
$<8\times 10^{-22}$ GeV$^{-5}$ &
$<7\times 10^{-21}$ GeV$^{-5}$ &
$<7\times 10^{-20}$ GeV$^{-5}$ &
$<2\times 10^{-20}$ GeV$^{-5}$
\\
\hline
$\big| \sum_{jm}\,_2Y_{jm} (\th,\ph) 
~\big(\kEdjm{4}{jm} + i\kBdjm{4}{jm}\big) \big|$ &
$\lesssim 10^{-37}$ &
$\lesssim 10^{-38}$ &
$\lesssim 10^{-38}$ &
$\lesssim 10^{-38}$
\\
$\big| \sum_{jm}\,_2Y_{jm} (\th,\ph) 
~\big(\kEdjm{6}{jm} + i\kBdjm{6}{jm}\big) \big|$ &
$\lesssim 10^{-31}$ GeV$^{-2}$ &
$\lesssim 10^{-32}$ GeV$^{-2}$ &
$\lesssim 10^{-31}$ GeV$^{-2}$ &
$\lesssim 10^{-31}$ GeV$^{-2}$
\\
$\big| \sum_{jm}\,_2Y_{jm} (\th,\ph) 
~\big(\kEdjm{8}{jm} + i\kBdjm{8}{jm}\big) \big|$ &
$\lesssim 10^{-25}$ GeV$^{-4}$ &
$\lesssim 10^{-25}$ GeV$^{-4}$ &
$\lesssim 10^{-24}$ GeV$^{-4}$ &
$\lesssim 10^{-24}$ GeV$^{-4}$
\end{tabular}
\caption{\label{bounds}
New constraints on coefficients for Lorentz and CPT violation.
The first three rows give constraints 
on general coefficients for CPT-odd operators with $d=5,7,9$.
The next three rows display constraints within the isotropic limit.
The final three rows provide the approximate maximal sensitivity 
of each source to coefficients for CPT-even operators with $d=4,6,8$.
}
\end{table*}

The CPT-even case is more complicated
because the normal modes are linearly polarized,
which implies linearly polarized light from a distant source
produced near one of the two polarizations 
$\vec s \approx \pm(\vs^1,\vs^2,0)$
could propagate essentially unchanged.
Moreover,
even if both modes are involved,
the change in polarization involves more than a simple rotation
of linear polarization.
Light that is initially linearly polarized,
$\vec s=(s^1,s^2,0)$,
becomes elliptically polarized as
$\vec s$ rotates out of the $s^1$-$s^2$ plane
and in some cases may even become circularly polarized, 
$\vec s = (0,0,s^3)$.

For light not produced in a normal mode
and initially linearly polarized at angle $\ps_0$,
let $\Psi = \ps_0-\ps_b$
be the difference between $\ps_0$ 
and the polarization angle $\ps_b$ for the faster of the two normal modes.
As the Stokes vector rotates about $\vec\vs$,
it traces out a cone with opening angle $4\Psi$
centered around $\vec\vs$.
A calculation shows that
the difference $\ps_2 - \ps_1$ 
in linear polarization 
at two different energies $E_1$ and $E_2$ satisfies
\bea
\sin 2(\ps_2-\ps_1) &= \fr{\sin2\Ps\cos2\Ps(\cos\Ph_2-\cos\Ph_1)}
      {\sqrt{(1-\sin^22\Ps \sin^2\Ph_1)(1-\sin^22\Ps \sin^2\Ph_2)}} ,
      \notag \\
\cos 2(\ps_2-\ps_1) &= \fr{\cos^22\Ps + \sin^22\Ps \cos\Ph_2\cos\Ph_1}
      {\sqrt{(1-\sin^22\Ps \sin^2\Ph_1)(1-\sin^22\Ps \sin^2\Ph_2)}} .
\eea
This result could be used to place constraints
when the observed difference $\ps_2-\ps_1$ is small.

An alternative and simpler strategy is again provided 
by considering the effective degree of linear polarization.
In the CPT-even case,
the effective degree of linear polarization is decreased
both by polarization smearing 
and by the conversion from linear to elliptical polarization.
The maximum effective degree of linear polarization is 
\beq
\Pi_\text{eff} = \sqrt{1- \big(1-\vev{\cos\Ph}^2\big) \sin^22\Ps} \ .
\eeq
An observation of linear polarization $\Pi$ 
then places a lower limit on $\Pi_\text{eff}$,
which leads to the inequality
$1-\Pi^2 > (1-\vev{\cos\Ph}^2)\sin^22\Ps$.
A single source therefore bounds a region in coefficient space
but cannot provide a strict constraint,
as its light could be propagating in a normal mode with $\Ps = 0$ or $\pi/2$.
In principle,
combining the results of multiple sources 
at different sky locations and having different polarizations
would permit a complete coverage of the CPT-even sector for each $d$,
but at present the number of sources is insufficient for this.
The above inequality can,
however, 
be used to estimate the maximum sensitivity 
to coefficients for Lorentz violation achieved by a given source.
The depletion in polarization is largest when the light is in
an equal admixture of the two normal modes,
$\Ps = \pm \pi/4$.
The effective degree of polarization then reduces to
$\Pi_\text{eff} = \big|\vev{\cos\Ph}\big|$,
which vanishes for relative rotations greater than $\pi$
across an energy band.
We can therefore estimate the maximal sensitivity to
the CPT-even coefficients as
\beq
\bigg|\sum_{jm}\syjm{2}{jm}(\th,\ph)
\big(\kE + i\kB\big)\bigg|
\lsim 
\fr{\pi}{2 \big|E_2^{d-3}-E_1^{d-3}\big| L^{(d)}} .
\quad
\eeq
This result can be applied to the sources
GRB 041219A, GRB 100826A, GRB 110301A, and GRB 110721A.
For Lorentz-violating operators of dimensions $d=4$, $6$, and $8$,
there are 10, 42, and 90 independent vacuum coefficients,
respectively
\cite{km09}.
Table \ref{bounds} lists the resulting constraints
on a linear combination of these coefficients
generated by each GRB.
They represent sensitivities improved by factors of 10 to 100,000
over existing bounds on CPT-even violations of Lorentz invariance.

This work was supported in part
by the Department of Energy
under grant DE-FG02-91ER40661
and by the Indiana University Center for Spacetime Symmetries.

\end{document}